\documentclass[twocolumn,prl,aps,epsf, reprint]{revtex4-1}
\usepackage{graphicx}
\usepackage{dcolumn}
\usepackage{bm}
\usepackage{hyperref}
\usepackage{mathtools}
\usepackage{booktabs}
\usepackage{xcolor}

\usepackage{titlesec}

\titleformat{\section}
{\normalfont\large\bfseries}{\thesection}{1em}{}

\begin{document}

\title{Anomalous Electronic Transport in High Mobility Corbino Rings }

\author{Sujatha Vijayakrishnan$^{1}$, F. Poitevin$^{1}$, Oulin Yu$^{1}$, Z. Berkson-Korenberg$^{1}$, M. Petrescu$^{1}$,  M.P Lilly$^{2}$, T. Szkopek$^{3}$, Kartiek Agarwal$^{1}$,  K. W. West$^{4}$, L. N. Pfeiffer$^{4}$  and G. Gervais$^{*1}$}

\affiliation{$^{1}$ Department of Physics, McGill University, Montr\'eal,  Qu\'ebec H3A 2T8 CANADA}

\affiliation{$^{2}$ Center for Integrated Nanotechnologies, Sandia National Laboratories, Albuquerque, New Mexico 87185, USA}

\affiliation{$^{3}$Department of Electrical and Computer Engineering, McGill University, Montr\'eal, Québec H3A 0E9 CANADA}

\affiliation{$^{4}$ Department of Electrical Engineering, Princeton University, Princeton, NJ  08544 USA}

\date{\today }

\begin{abstract}
We report low-temperature electronic transport measurements performed in two multi-terminal Corbino samples formed in GaAs/Al-GaAs two-dimensional electron gases (2DEG) with both ultra-high electron mobility ($\gtrsim 20\times 10^6$ $cm^2/Vs)$ and with distinct electron density of $1.7$ and $3.6\times 10^{11}~cm^{-2}$. In both Corbino samples, a non-monotonic behavior  is observed in the temperature dependence of the resistance below 1~$K$. Surprisingly, a sharp decrease in resistance is observed with increasing temperature in the sample with lower electron density, whereas an opposite behavior is observed in the sample with higher density. To investigate further, transport measurements were  performed in large van der Pauw samples having identical heterostructures, and as expected they exhibit resistivity that is monotonic with temperature. Finally, we discuss the results in terms of various lengthscales leading to ballistic and hydrodynamic electronic transport, as well as a possible Gurzhi effect.

 \end{abstract}

\maketitle 
\section{Introduction}
Over the last two decades, great progress has been achieved in increasing the electron mobility in two-dimensional electron gases  formed in MBE-grown materials such as GaA/AlGaAs and alternatively in exfoliated graphene. Spectacularly, the electron mobility in GaAs/AlGaAs 2DEGs has recently been reported  to reach $57\times 10^6$ $cm^2 (Vs)^{-1}$ \cite{Chung2022} and, in the absence of phonons at low temperatures, this results in large impurity-dominated mean free path that can exceed $350 ~\mu m$.  These high-mobility 2DEGs  are notoriously well described by Fermi liquid theory at low temperatures, but what is perhaps less obvious is that counter-intuitive phenomena can arise as a result of an interplay between hydrodynamic transport and confinement.  As an example, Gurzhi noted  in 1963  \cite{Gurzhi1963,Gurzhi1968} that if a  Fermi liquid is confined in a narrow constriction of characteristic size $d$, within some restrictive conditions the resistance of the metal could decrease with increasing temperature. More recently, studying the scattering lengths in 2D semiconductors with moderately high electron mobility,  Ahn and Das Sarma \cite{Sarma2022} proposed that Gurzhi's prediction could occur even in bulk GaAs 2DEGs with sufficiently short electron-electron scattering lengths and low disorder. \\ 
 
Motivated by these works, we have fabricated two identical multi-terminal Corbino rings in GaAs/AlGaAs 2DEGs with electron mobility exceeding $20\times 10^6$ $cm^2 (Vs)^{-1}$, and with two different electron density leading to distinctive electron-electron and electron-impurity scattering lengths. Four-point conductance (resistance) measurements were performed in these Corbino with a transport channel defined by a $40~\mu m$  annular ring  probing only the bulk of the sample, {\it i.e.} with no edge. These measurements are compared with similar measurements performed in millimetre scale van der Pauw (VdP) samples that have the same heterostructure. We note the intrinsic resistivity (conductivity) in the VdP and Corbino samples differs from the measured resistance (conductance) solely by a geometric factor and therefore both will be used interchangeably in the text below. Astonishingly, the temperature dependence of both multi-terminal Corbino shows an anomalous temperature dependence whereby in one case a sharp decrease in resistance is observed at temperatures below 1~$K$ with increasing temperature, and an opposite behavior is observed in the other Corbino sample. In the  van der Pauw (VdP) samples with identical heterostructure to the multi-terminal Corbino, as expected a monotonic behavior of resistivity with temperature was observed.\\

\section{Results}

\begin{figure}[!t]
  \centering
      \includegraphics[width=\columnwidth]{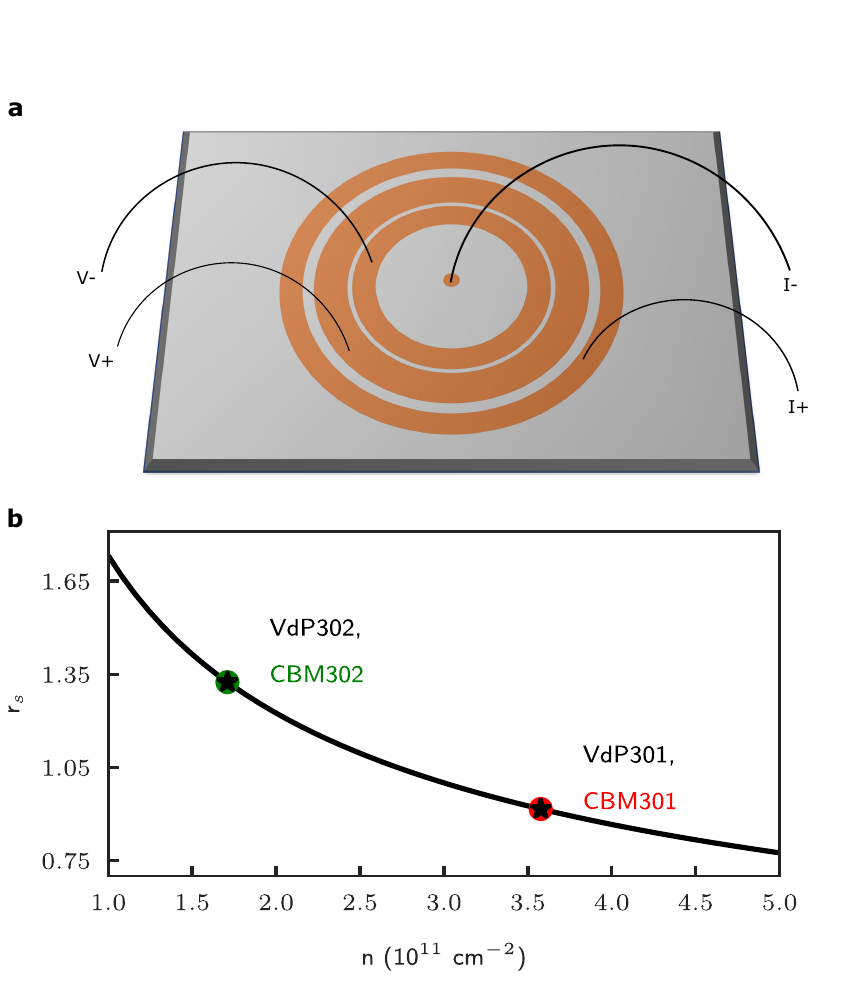}
  \caption {\textbf{Schematics of the multi-terminal Corbino and electron-electron interaction parameter.} ({\bf a}) Four-terminal Corbino design with the gold rings depicting the contact geometry for sample CBM301 and CBM302. The contact ring sizes is identical for both samples. ({\bf b}) Electron-electron interaction parameter $r_s$ {\it versus } electron density (see main text).  The $r_s$ parameter values for CBM301 (and VdP301) denoted as red dot (star) and CBM302 (and VdP302) denoted as green dot (star) are 0.92 and 1.32, respectively.}
\label{fig:Fig1}
\end{figure}

{\bf Corbino and van der Pauw geometry.} Typical measurements performed in large 2DEG samples usually utilizes either samples prepared in van der Pauw or Hall bar geometry. In the VdP geometry, the electrical contacts are located in each corner and midpoint along the perimeter of a  square wafer of size $3\times3$ mm in our case. Four-point measurements are then performed  by applying a fixed current (voltage) and measuring voltage (current) using a combination of four contacts. Even though non-patterning helps to preserve the pristine state of the ultra-high quality 2DEG, transport measurements performed in either VdP or Hall bar unavoidably contain both the bulk and surface (or more precisely the edge) effects.  In the case where only the bulk contribution to the resistivity (or conductivity) is wanted, one can have recourse to the Corbino geometry where the sample contacts are prepared in a circular/ring geometry, with the active region of the 2DEG sample defined by an annulus. In this case, the current is applied concentrically from the outer to inner radii and the electronic transport measurements solely probe the bulk of the sample. Most studies performed so far in the Corbino geometry have focused on samples with an inner contact and a single outer ring and these are inherently two-point contact measurements where unfortunately the contact resistance with the 2DEG is not eliminated. However, multi-terminal Corbino samples possessing one inner contact and three outer rings can be fabricated for four-point measurements, see FIG.\ref{fig:Fig1}(a). While the Corbino geometry does not have any edge nor radial dimension, however it does have a channel length which is defined as the distance between the inner diameter of the third ring ($V_+$ probe) and the outer diameter of the second ring ($V_-$ probe). In our case, this channel length is $L_{Corbino}=40~ \mu m$ for both Corbino samples. \\

{\bf Electron density and $r_s$ parameter.}  The dimensionless electron-electron interaction parameter $r_{s}=(\pi n a_{B}^{2})^{- \frac{1}{2}}$, where $n$ is the electron density of either set of samples, and $a_B = 103~ \AA $ is the effective Bohr magneton radius, are 0.92 and 1.32 for the 301 and 302 wafers, respectively, and they are shown in FIG.\ref{fig:Fig1}(b). Given its significantly higher $r_s$ value and ultra-high electron mobility,  the 2DEG formed in 302 samples (when compared to 301) are expected to inherently have enhanced electron-electron interaction, and conversely significantly distinct electron-electron scattering lengths versus temperature.  \\

\begin{figure}[!t]
  \centering
      \includegraphics[width=\columnwidth]{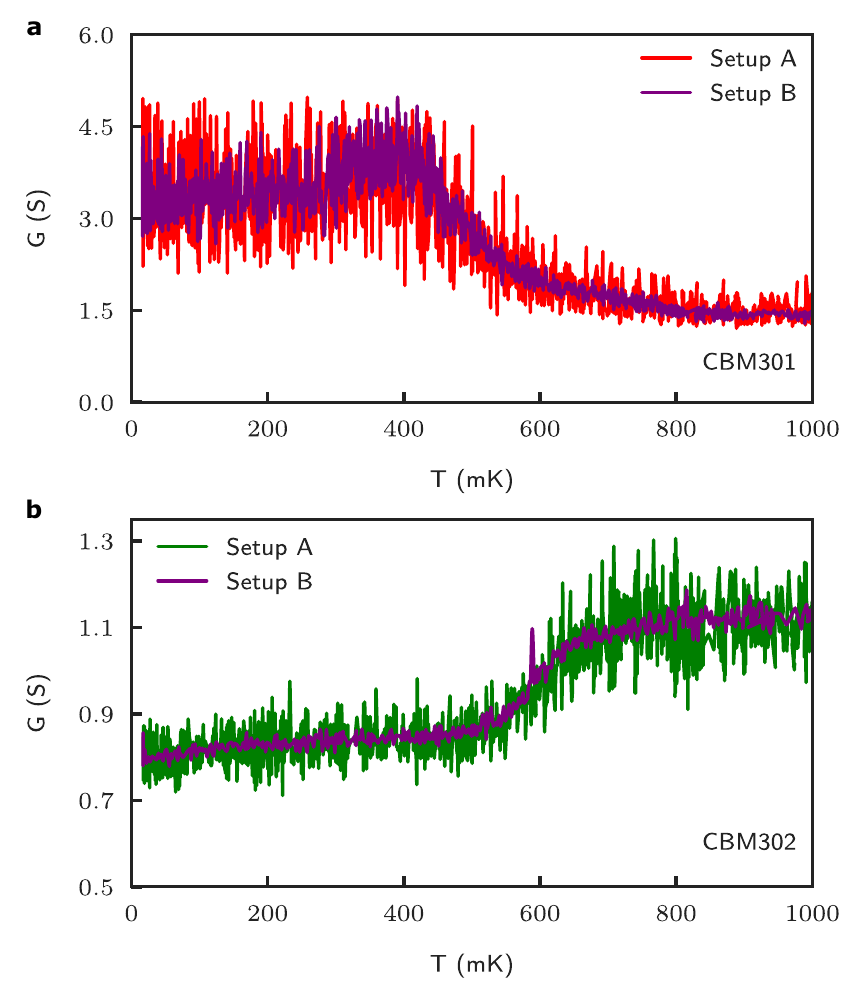}
\caption{\textbf{Temperature dependence of the Corbino conductance.} The four-point conductance $G$ measured with two distinct measurement setups labeled A and B (see main text and SM),  is shown $\it{versus}$ temperature  for  CBM301 ({\bf a}) and CBM302 ({\bf b}). An increase in conductance (decrease in resistance) onsetting near 500 $mK$ is observed in CBM302 (panel b) with increasing temperature, whereas an opposite behavior at $\sim$400 $mK$ is observed for CBM301 (panel a).  Note that in the next figures the resistance $R=G^{-1}$ rather than the conductance will be shown. }
\label{fig:Fig2}
\end{figure}

\begin{figure}
  \centering
      \includegraphics[width=\columnwidth]{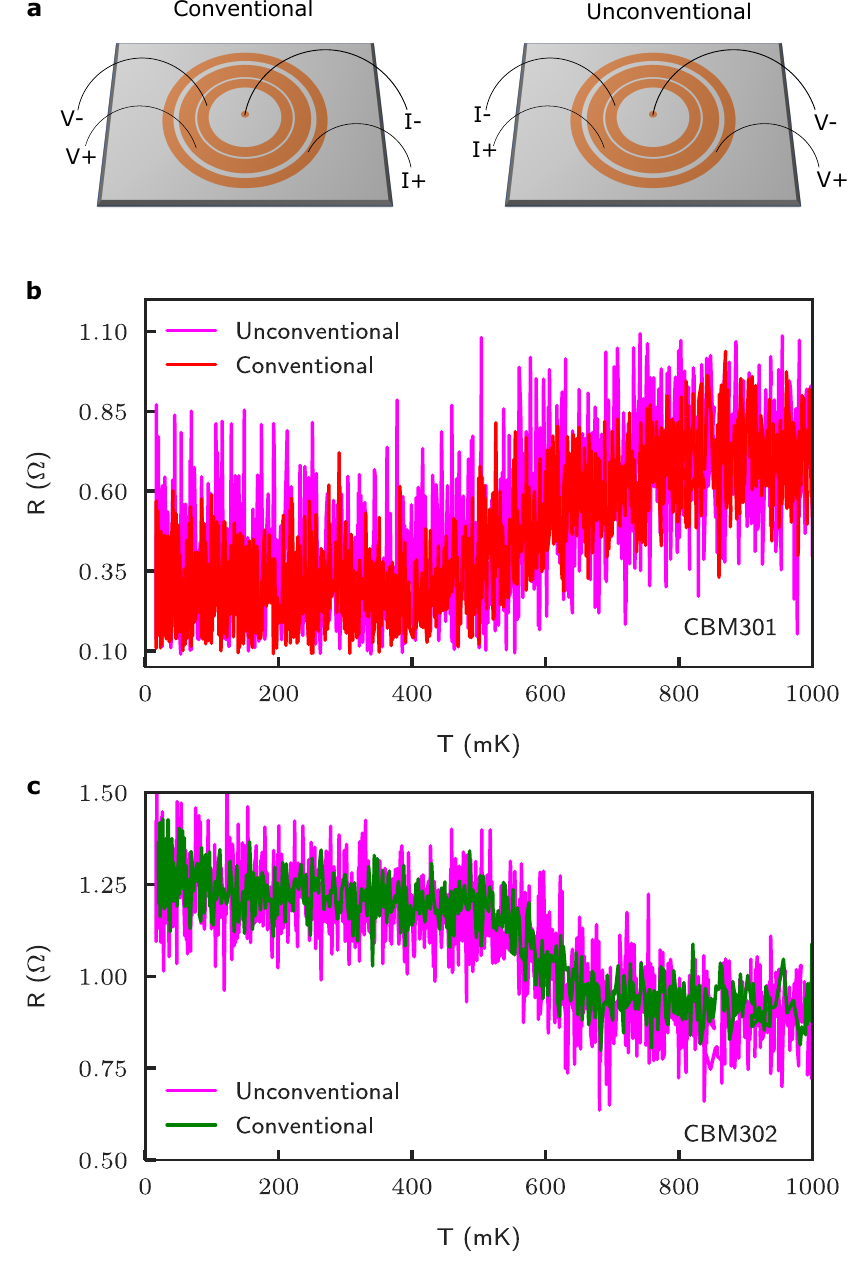}
  \caption{\textbf{Conventional and unconventional resistance measurement.} ({\bf a}) Schematics of the current-voltage probe terminal used for conventional (left) and unconventional (right) measurements.  In the unconventional case, the current ($I_+, I_-$) and voltage ($V_+, V_-$) leads are interchanged, with the current leads connected to the inner Corbino rings. The resistance measured  {\it versus} temperature for CBM301 ({\bf b})  and CBM302 ({\bf c}) is shown for both cases, with the data in magenta denoting the unconventional configuration. The reciprocity relations are clearly validated in both Corbino samples.}
\label{fig:Fig3}
\end{figure}

{\bf Four-point Corbino conductance measurements.} Electronic transport measurements were carried on with the multi-terminal Corbino samples. In this geometry, it is usual to measure the conductance and for this reason we first show the measured conductance {\it versus} temperatures in both samples. We performed these measurements with  two distinct measurement circuits, labeled setup A (20 $nA$ excitation current) and setup B (83 $nA$ excitation current), see supplementary material (SM). These results are shown in  FIG.\ref{fig:Fig2}(a) for CBM301 and FIG.\ref{fig:Fig2}(b)  for CBM302, and the data obtained with each circuit are in excellent agreement with one another. The temperature dependence of conductance measured in CBM301 shows a decrease in conductance at temperatures above $\sim$400 $mK$, which at first sight may not be surprising given there has been reports of increasing electron mobility well below 1 $K$ in 2DEGs of moderately high mobility\cite{Umansky1997}. Surprisingly, a completely opposite trend is observed for CBM302 where an anomalous increase in conductance (decrease in resistance) with increasing temperature is observed at temperatures above $\sim$500 $mK$. The resulting fractional change in conductance ($\Delta G/G$) is $\sim40\%$ between $\sim$500 and 700 $mK$, and this is different from the anomalous transport behavior observed in CBM301 in spite of both samples having an identical geometry and similar high electron mobility.\\

\begin{figure}[!t]
  \centering
       \includegraphics[width=\columnwidth]{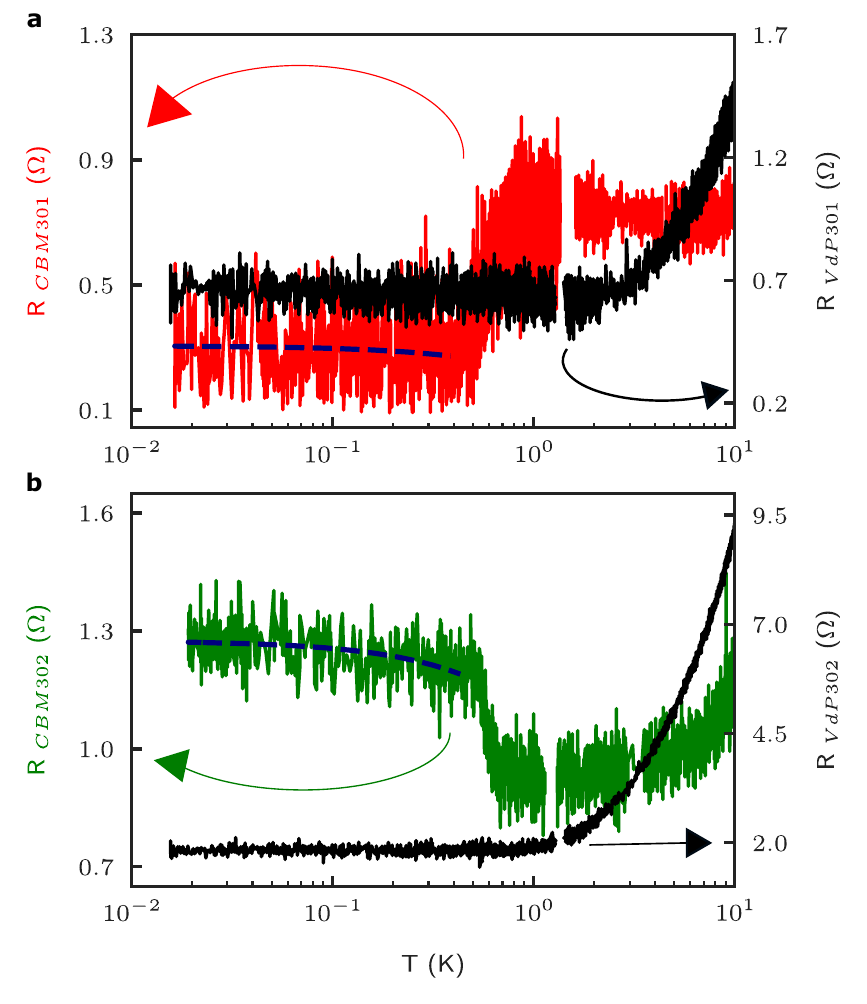}
  \caption{\textbf{Temperature dependence of Corbino and van der Pauw resistance (VdP).} The temperature dependence of the measured Corbino resistance is compared with the resistance measured in VdP cut from wafers with the exact same heterostructure. Panel ({\bf a}) shows resistance measured for CBM301 (red) and VdP301 (black). Panel ({\bf b}) shows the resistance measured for CBM302 (green), and VdP302 (black). A guide-to-the-eye blue dash line shows the moving average of the resistance at low temperature for CBM301 and CBM302.} 
\label{fig:Fig4}
\end{figure}

{\bf Probe symmetry and reciprocity theorem.} To further validate our observation, we have performed conventional and unconventional transport measurements in both Corbino samples, see FIG.\ref{fig:Fig3}. Note that in this figure, and in the remainder of the text we opted to report and discuss the resistance $R=G^{-1}$ {\it versus} temperature rather than the conductance. In the unconventional configuration, the current ($I_+, I_-$) and voltage ($V_+, V_-$) probes are interchanged and this results in the Corbino geometry for the current probes, somewhat counterintuitively, to be connected to intermediate rings, and the voltage probes to the most outer and inner contacts, see FIG.\ref{fig:Fig3}(a). The reciprocity theorem of electromagnetism which traces its roots to Maxwell's equations (and has been connected to Onsager's relations by Casimir \cite{Casimir1963}) states that both configurations should yield the same measurement. This holds true for passive circuits that are composed of linear media and for which time-reversal symmetry is not broken. Our reciprocity measurements are shown in FIG.\ref{fig:Fig3}(b) and (c), with the data shown in magenta taken in the unconventional configuration. Except for the higher noise floor, the measurements clearly satisfy the reciprocity theorem, adding strength to the observation of anomalous transport in the multi-terminal Corbino samples.\\

{\bf Comparison of Corbino and van der Pauw measurements.}  van der Pauw measurements were performed to determine the bulk resistivity of each parent heterostructure. Fig.\ref{fig:Fig4}(a) and (b) shows the resistance measured in both Corbino and VdP samples up to 10 $K$ temperature on a semi-log scale. The resistance of VdP301 shows a very slight decrease up to 2 $K$ which is then followed by the expected monotonic increase due to scattering with phonons. This behavior is drastically distinct from CBM301 which exhibits a sudden increase in resistance at $\sim$400 $mK$,  followed by a saturation over a wide range of temperatures, up to $\sim$10 $K$. In the case of VdP302, the increase  in resistance due to phonons onsets near 1 $K$, and as expected is followed by a monotonic increase with temperature, see Fig.\ref{fig:Fig4}(b).  This greatly contrasts with the sudden decrease in resistance observed in CBM302 near $\sim$500 $mK$, followed by a saturation up to a temperature of $\sim$3 $K$ where a monotonic increase with temperature is observed. We also note that in all cases the resistance is nearly constant at temperatures below $\sim$0.5 $K$, as expected in very high-mobility GaAs/AlGaAs 2DEGs, hence ruling out the anomalous behaviour in electronic transport observed being caused by a percolation process (metal-insulator transition). Recent work also found that under some conditions NiGeAu used to create the electrical contacts could be superconducting with a transition temperature around $\sim$700~$mK$ and with a critical field of $\sim$0.15~$T$ \cite{Ritchie2020,Saunders2022}. In order to investigate if a supercurrent and/or a proximity effect could have played a role in our transport measurements, two-terminal and four-terminal differential resistance measurements under a DC current bias were performed  at base temperature of the dilution refrigerator ($T\sim 20~ mK$) and the $I-V$ obtained were found to be linear (Ohmic). In addition, we have performed magneto-transport measurements in a perpendicular field near the expected critical field $0.15~T$ (also at base temperature) and found no evidence of superconductivity in either samples, see SM. Finally, additional electronic transport data up to 100 $K$ in the Corbino and up to 10 $K$  in the VdP samples are provided in the SM that illustrates the distinctive behaviour between the VdP and CBM samples at temperatures below a few Kelvins.\\

\section{Discussion}

{\bf Transport lengthscales.}  As discussed in Ref.  \cite{Sarma2022}, the relevant lengthscales that are impacting the bulk resistivity of a high-mobility 2DEG are: (A) The momentum-conserving mean free path due to electron-electron interaction, $l_{ee}$; (B) the momentum-relaxing mean free path due to electron scattering with both impurities and phonons, $l_e$;  and (C) the width of the annular region forming the transport channel which is $L_{Corbino}=40~\mu m$ in our case.  When phonons contribute negligibly to electron scattering, which has been shown to be the case below $1~K$ temperature \cite{Sarma2008}, the momentum-relaxing mean free path $l_e$ can be calculated from the mobility and electron density,  $l_{e}=5.22 \times \mu \sqrt{n}$. Owing to the very high-mobility of the 2DEG used here, this lengthscale is exceedingly larger than the channel transport length, yielding  $\sim$271 $\mu m$ in CBM301 and  $\sim$138~$\mu m$ in CBM302.  \\

Within Fermi liquid theory, Ahn and Das Sarma \cite{Sarma2022} calculated from first principles the momentum-conserving mean free path $l_{ee}$ for 2DEGs with similar density and with an electron mobility of $1\times 10^{6}$  $cm^2 (V~s)^{-1}$ . In particular, they found that for a 2DEG with a density $1.5\times 10^{11}$  $cm^{-2}$ (similar to CBM302) the momentum-conserving mean free path $l_{ee}$  falls below 40 $\mu m$ at temperatures above $\sim$ 550 $mK$.  At a higher electron density of $2.5 \times 10^{11}$  $cm^{-2}$ (close to CBM301), this occurs at  a temperature of $\sim$850 $mK$ and so the condition $l_{ee} < L_{Corbino} \ll l_e $ can occur within a finite temperature  interval for both CBM301 and CBM302 with a negligible phonon contribution to resistivity, {\it i.e.} until $l_e$ is reduced due to phonon scattering and becomes comparable to either $l_{ee}$ or $L_{Corbino}$. \\

{\bf Knudsen parameter, ballistic transport and hydrodynamics.} In fluids, when a liquid or a gas flow through an aperture, distinctive flow regime can be classified by a dimensionless parameter known as the Knudsen number $K_n=\lambda/D$. Here,  $\lambda$ is the mean free path of the fluid's constituent and $D$ is the diameter of the aperture or fluid channel. As a function of $K_n$, a classical fluid can transit from (effusive) single-particle  transport described by kinetic theory of statistical mechanics at high Knudsen number ($K_n\gg 1$), to a continuous hydrodynamic flow governed by the Navier-Stokes at low Knudsen number ($K_n\ll 1$), see Ref.\cite{Savard2009} for a simple experimental demonstration in the case of a gas flowing through a small aperture. \\

In electronic systems such as two-dimensional electron gases, both transport regimes are in principle possible although the realization of hydrodynamic flow has proven to be challenging to observe experimentally. In their study of transport-relevant lengthscales in high-mobility 2DEGs, Ahn and Das Sarma \cite{Sarma2022} have defined a dimensionless  Knudsen parameter, $\zeta \equiv  l_{ee}/l_e$, which also marks the  crossover from ballistic (or effusive) flow  when $\zeta \gg 1$, to the hydrodynamic regime in a continuum  when $\zeta \ll 1$. Interestingly, this $\zeta$ parameter falls below 0.5 at temperatures above 500 $mK$ for both CBM301 and CBM302, hinting strongly to the occurrence of hydrodynamic transport in a regime where phonons play very little, or no role at all. We note that the sharp drop in resistance at $\sim$500 $mK$ observed in CBM302  occurs when  $l_{ee} < L_{Corbino}$ and  $\zeta \ll 1$, suggestive of a hydrodynamic flow caused by increased electron-electron interaction. To our knowledge, this sharp drop (increase) observed in CBM302 (CBM301) in resistivity with increasing temperature below 1 $K$ has never been observed before in any high-mobility 2DEGs. Moreover, this is  surprising given the transport channel length in the Corbino being 40 $\mu m$ long, and for which a priori one would assume to be well within the bulk regime of 2D electronic flow.\\

{\bf Gurzhi effect?} The prediction by Gurzhi that a Fermi liquid metal can have a decreasing resistance with increasing temperature due to hydrodynamic flow has been notoriously difficult to realize in 3D metals, with the end result that very little progress has been made on the topic for decades. Only over the last few years there has been experimental report of hydrodynamic behaviour in 3D materials \cite{Moll2016,Gooth2018}. In 2D semiconductors, materials have long been made with astonishingly high-electron mobility and hence extremely long mean free path. In spite of this, reports of hydrodynamic flow have been scarce,  with the first claim made roughly thirty years ago (and only one for many years) in GaAs quantum wires \cite{Jong1995}. More recent works have focused on bilayers \cite{Gusev2021}, strongly-correlated 2D hole systems \cite{Kumar2021arXiv}, non-local measurements \cite{Gupta2021,Gusev2018} including an extensive study of the conductivity in GaAs/AlGaAs channel with smooth sidewalls and perfect slip boundary condition \cite{Keser2021}, graphene \cite{Kumar2017,Bandurin2018,Ku2020,Alec2022}, graphene Corbino rings \cite{Kumar2021} as well as theoretical interest of hydrodynamic flow in Corbino geometries \cite{Falkovich2019,Andreev2022,Andreev2022N,Narozhny2022}. But to the best of our knowledge, there has been no report to date of a direct observation of Gurzhi's prediction of a  super ballistic-to-hydrodynamic electron flow when phonons play very little, or no role at all.\\   

The Gurzhi effect occurs when the electron flow is neither diffused by impurity scattering nor  by single-particle transport events but is rather the result of the formation of a continuum and collective transport properties described by hydrodynamics.  In principle, this effect can be observed in the resistivity of a clean Fermi liquid metal up to a temperature where  the increasing electron-phonon interaction becomes important and leads to a non-negligible scattering sources. Rather stunningly, the resistance observed below $\sim1$ $K$ in CBM302 shows a great similarity  with the resistance curve predicted and plotted by Gurzhi in his original work \cite{Gurzhi1963}. This possibility  is further supported by the $\zeta$ value calculated which would indicate the electron flow to be in the hydrodynamic regime. This being said, in the case of CBM301 whose higher electron density leads to a smaller $r_s$ parameter and hence with decreased electron interaction when compared to CBM302, the sudden increase in resistance around 400 $mK$ with temperature cannot be explained solely by arguments based on hydrodynamic flow, nor by  any obvious electron-phonon scattering mechanisms.\\

A recent study conducted on spatial mapping of local electron density fluctuation in a high-mobility GaAs/AlGaAs 2DEG by way of scanning photoluminescence  \cite{Pfeiffer2019} reported electron density variations up to 100 $\mu m$ with a spot size of 40 $\mu m$. These fluctuations are likely to generate local electron mobility fluctuations, and 
we hypothesize that it could perhaps play a role in a Corbino measurement scheme because there is no edge and the concentric sample  can be viewed as a very large number of conductors wired in parallel.  This being said, whether a higher conductance path due to a local spatial fluctuation in the electron density (or mobility) would occur, and lead to the anomalous electronic transport observed in both CBM301 and CBM302 remains an open question. This will be the subject of future works.\\

To summarize, we have performed four-terminal electronic transport measurements in two very-high mobility Corbino 2DEG rings with distinct electron density and identical annular channel length of 40 $\mu m$. In both cases, anomalous transport was observed in the temperature dependence of the resistance at temperatures below 1 $K$ where phonons are expected to play a negligible or no role at all. The discovery of a sharp decrease in resistance with increasing temperature in the lower density sample is stunning, and even more so since the estimated Knudsen parameter hints at a Gurzhi effect and a crossover from super-ballistic to hydrodynamic flow. Nevertheless, this sharp decrease in resistance contrasts with the trend observed at a similar temperature in the higher electron density sample, and for which we have no clear explanation as for its origin. At temperatures above 10 $K$, we have shown that in both cases  the expected monotonic temperature dependence of the resistance is recovered and is similar in trend to that measured in larger millimetre size VdP samples.  While the exact mechanism leading to the anomalous electronic transport observed here remains an open question, this work demonstrates that a 40 $\mu m$ channel length is  not bulk in high mobility 2DEGs since both the momentum-conserving and momentum-relaxing mean free path values are either larger,  or equal to the channel length in the 20 $mK$  to $\sim$1 $K$ temperature range. Looking forward, confirmation of hydrodynamic flow in GaAs/AlGaAs could allow us to study the remarkable properties of the anti-symmetric part of a viscosity tensor, known as the Hall viscosity  \cite{Avron1995,Tobias2019}, a dissipationless viscosity existing even at zero temperature that has no classical equivalent whatsoever.\\

\section{Methods}

{\bf Experimental design and procedure.} The measurements were performed on GaAs/AlGaAs symmetrically-doped heterostructures with a quantum well width of 30 $nm$ (CBM301, VdP301) and 40 $nm$ (CBM302, VdP302. A sketch of the heterostructure that include the main growth parameters such as the setback distances for the dopants and the capping layer thickness is provided in the SM for both samples. The electron density  of CBM301 and VdP301 is $3.6 \times 10^{11}$ $cm^{-2}$, and for  CBM302 and VdP302 is $1.7\times 10^{11}$ $cm^{-2}$, as determined by magneto-transport measurements of Shubnikov de-Haas (SdH) oscillations at low magnetic fields, see SM. Their mobilities are $27.8 \times 10^{6}$ $cm^2 (Vs)^{-1}$ and $20.3 \times 10^{6}$ $cm^2 (Vs)^{-1}$, respectively.\\

The electrical contacts for the Corbino samples were patterned using UV lithography followed by e-beam deposition of Ge/Au/Ni/Au layers of $26/54/14/100$ ~$nm$  thickness and at a $1-2 ~A~s^{-1}$ rate. Subsequently, the contacts were annealed in H2N2 atmosphere using a two step annealing procedure: a $20 ~s$ first step at $370 ~^{\circ}C$ followed by a longer $80 ~s$ step at $440 ~^{\circ}C$ \cite{Ben2015,Bennaceur2018}.\\

Both Corbino samples have three 2DEG rings with inner/outer radii $(150/750)~\mu m$, $(960/1000)~\mu m$, and $(1300/1400)~\mu m$, respectively, see FIG.\ref{fig:Fig1}(a). One van der Pauw sample was cleaved from the exact same wafer used to fabricate CBM302, and from a twin wafer with an identical heterostructure grown on the same day for CBM301. They are 3 $\times$ 3~mm square wafers with eight diffused indium contacts around the perimeter.  All samples were cooled to a base temperature of approximately 20 $mK$ in a dilution refrigerator, and were illuminated by a red LED from room temperature down to approximately 6 $K$ to increase both the mobility and density of the 2DEG. Specific details pertaining to the circuit used to perform the transport measurements are provided in the SM.\\

\section{Data availability}

The data presented in this work are available from the corresponding author upon reasonable request.\\

\section{Acknowledgments}

This work has been supported by NSERC (Canada), FRQNT-funded strategic clusters INTRIQ (Québec), PromptInnov/MEI PSR Quantique (Québec) in partnership with Mitacs and Montreal-based CXC. The work at Princeton University is funded in part by the Gordon and Betty Moore Foundation’s EPiQS Initiative, Grant GBMF9615 to L. N. Pfeiffer, and by the National Science Foundation MRSEC grant DMR 2011750 to Princeton University. Sample fabrication was carried out at the McGill Nanotools Microfabrication facility. We would like to thank B.A. Schmidt and K. Bennaceur for their technical expertise during the fabrication and the earlier characterization of the Corbino sample, and R. Talbot, R. Gagnon, and J. Smeros for general laboratory technical assistance.\\

\section{Authors contributions}

S.V. and G.G. conceived the experiment.  K.W.W. and L.N.P. performed the semiconductor growth by molecular beam epitaxy and provided the material. S.V. fabricated the Corbino and prepared the VdP samples. S.V. performed the electronic transport measurement  in the Corbino samples at low temperatures, with the assistance and expertise of F.P. and M.P. O.Y. assisted in noise reduction and pre-amplification. S.V. performed the data analysis, helped by  Z.B.K. for the development of computer routine. M.P.L. and T.S. provided important expertise regarding data acquisition, semiconductor expertise, and interpretation of the results. K.A. provided theoretical guidance.  S.V. and G.G, wrote the manuscript, and all authors commented on it.\\

{\bf Competing interests} : The authors declare no competing interests.\\

{\bf Corresponding author information} : Guillaume Gervais, gervais@physics.mcgill.ca.

 \nocite{*}
\bibliographystyle{apsrev4-1}

\pagebreak

\widetext
\newpage
\begin{center}
\textbf{\large Supplementary Material}
\end{center}

\setcounter{equation}{0}
\setcounter{figure}{0}
\setcounter{table}{0}
\setcounter{page}{1}
\makeatletter
\renewcommand{\theequation}{S\arabic{equation}}
\renewcommand{\thefigure}{S\arabic{figure}}
\renewcommand{\bibnumfmt}[1]{[S#1]}
\renewcommand{\citenumfont}[1]{S#1}
\section {Heterostructure}
A sketch of the wafer heterostructure used to fabricate CBM301, VdP301  and CBM302, VdP302  is shown in Fig.S1. The main components forming the heterostructure grown on an undoped GaAs subtrate are:  a buffer/spacer, the setback defining the position of the dopants, the quantum well width, and the spacer/capping layer. The main differences between the 301 and 302 
heterostructures  are: the electron density ($3.6\times 10^{11}~ cm^{-2}$ for CBM301 and $1.7 \times 10^{11}~ cm^{-2}$ for CBM302); the quantum well width (30~$nm$ for CBM301 and 40~$nm$ for CBM302);  the setback distance for the dopants ($\sim$80~$nm$ for CBM301 and $\sim$160~$nm$ for CBM302).\\

\begin{figure}[!ht]
	\centering
      \includegraphics[width=121mm]{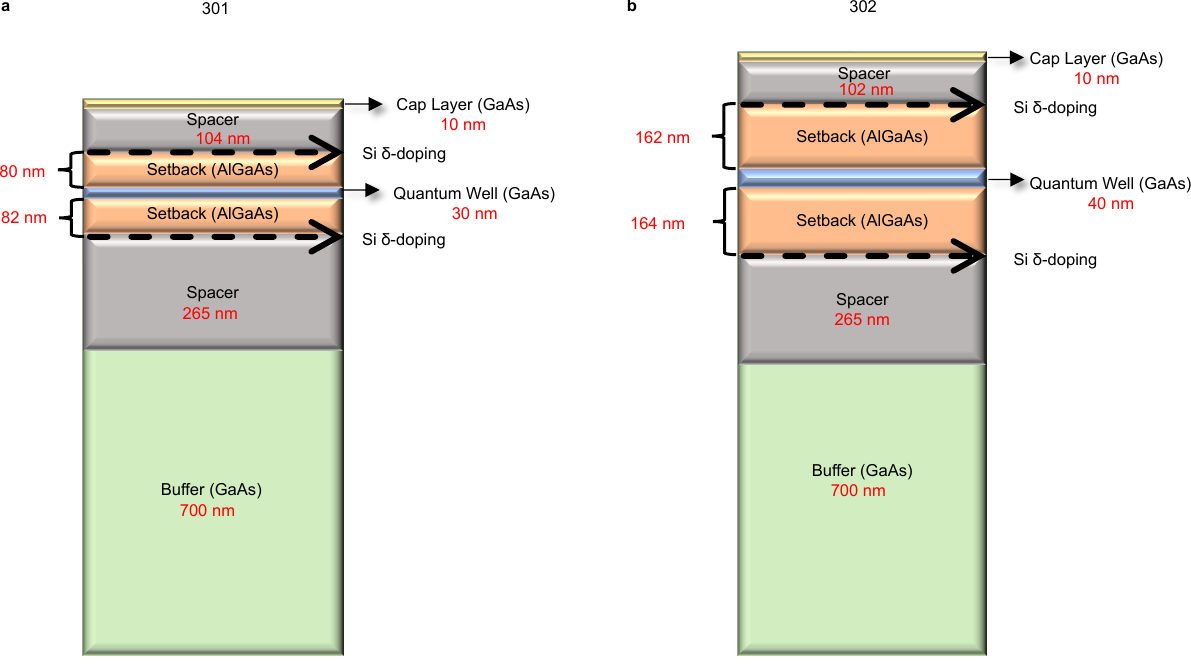}
  \caption{Heterostructure of (a) 301 (CBM301, VdP301)  and (b) 302 (CBM302, VdP302). The main heterostructure components and their thickness are shown.}
  \label{fig:FigS1}
\end{figure}

\section {Electronic transport measurement circuit}
In the main text of the manuscript, Fig.2 shows the conductance of Corbino devices measured with two different electrical setups, and these are discussed below.

\subsection{Experimental setup A}

The experimental setup A  shown in Fig.S2 consists of a SR830 lock-in amplifier and a resistor with a high resistance. An output voltage of 200 $mV$ is passed through  a 10 $M\Omega$ resistor connected in series with the Corbino device. This configuration allows us to apply a constant current of 20 $nA$ to the outermost contact of the Corbino while keeping the inner contact grounded. The four-point resistance is calculated from the voltage drop measured across the inner two rings of the Corbino samples.

\begin{figure}[!ht]
	\centering
      \includegraphics[width=121mm]{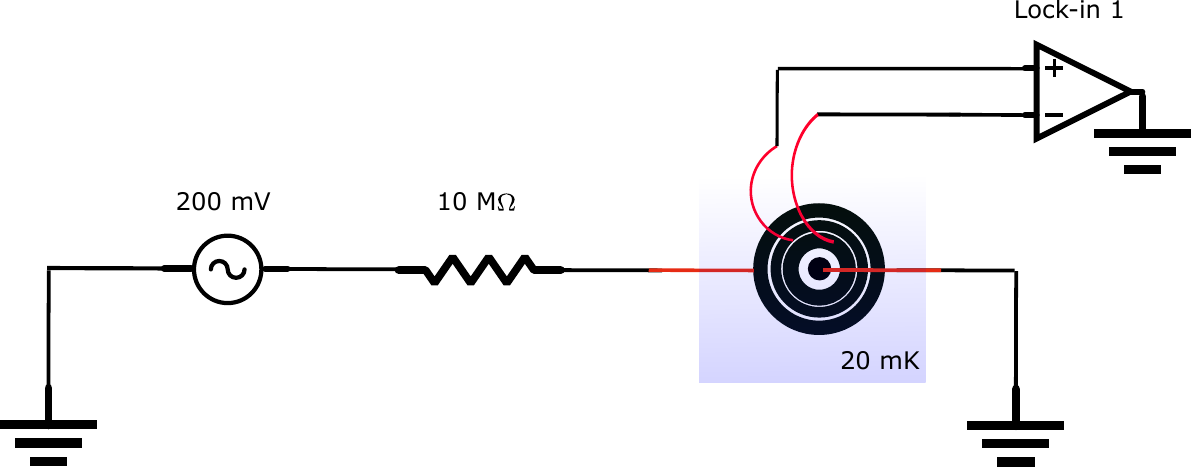}
  \caption{Setup A circuit used to determine the four point resistance of the Corbino sample with a fixed current of 20 $nA$.}
  \label{fig:FigS2}
\end{figure}

\subsection{Experimental setup B}
In the Corbino geometry, the sample has a large magneto-resistance in non-zero magnetic fields, which makes the previous experimental setup not ideal for this case. Experimental setup B  was used to measure the conductance while sweeping the magnetic field in order to extract the electron density from the Shubnikov de-Haas (SdH) oscillations. In this configuration, an output voltage of 100 $mV$  was applied to a voltage divider consisting of 100 $k\Omega$ $\parallel$ 100 $\Omega$, connected in series with the Corbino. The current through the sample is calculated from the measured voltage drop across a 1 $k\Omega$ resistor connected in series with the Corbino. For both experimental setups A and B, a voltage pre-amp with the gain of 100 was used at the output end.
\begin{figure}[!ht] 
	\centering
      \includegraphics[width=121mm]{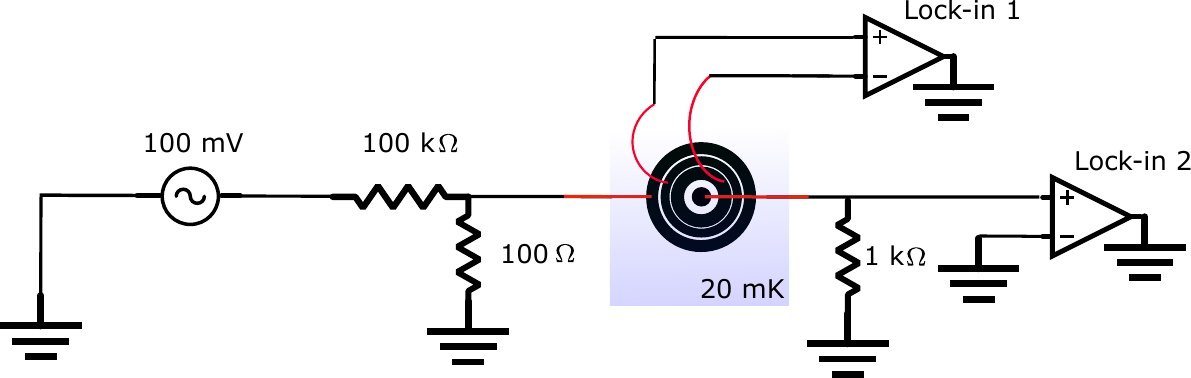}
  \caption{The circuit used to measure the conductance at B$\neq$ 0. This circuit was used in Fig.2 of the main manuscript to compare and verify the result obtained with setup A.}
   \label{fig:FigS3}
\end{figure}
\\

\newpage
\section{Magneto-transport measurement}
In  Fig.S4 the magneto-conductance measured at low perpendicular magnetic fields with the experimental setup B and is shown. The onset of Shubnikov-de Haas oscillations (SdH) is observed at very low magnetic fields as expected from the very high electron mobility of the 2DEGs. Due to a large magneto-conductance at low magnetic field in a Corbino sample, distinct panels show the conductance over different magnetic field ranges. No evidence was found for a superconducting transition, {\it i..e}  a superconducting critical field, in either samples.

\begin{figure}[!ht]
	\centering
      \includegraphics[width=\textwidth]{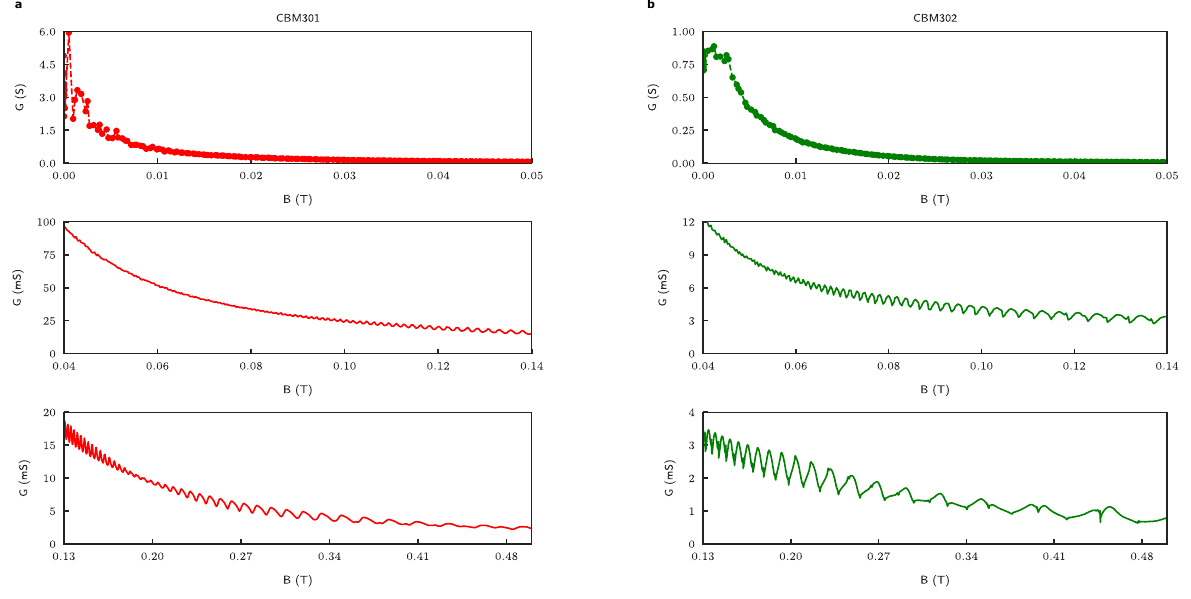}
     \caption{Magneto-conductance measured in (a) CBM301 and (b) CBM302 at 20 $mK$.}
     \label{fig:FigS4}
\end{figure}

\section {Electron density extraction }
\begin{figure}[!ht]
	\centering
      \includegraphics[width=121mm]{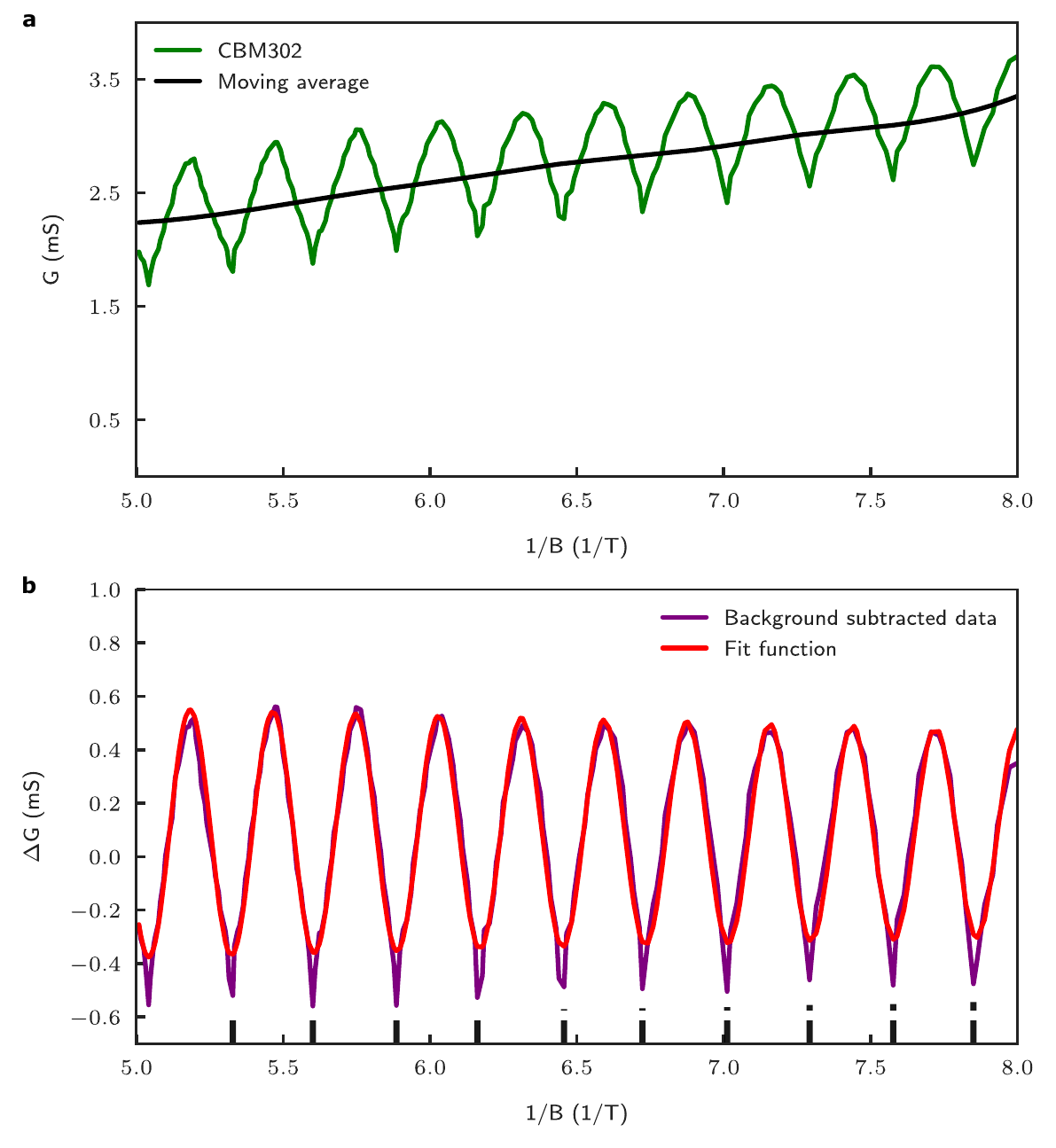}
  \caption{Panel (a) shows the SdH oscillation of CBM302 at 20 $mK$ and the calculated moving average. The background reduced data and fit performed are shown in panel (b).}
     \label{fig:FigS5}
\end{figure}

The SdH oscillations (after reducing the background) of conductivity (or conductance) {\it versus} inverse magnetic field are given by,
\begin{equation}
\Delta \sigma_{xx} \propto A~\cos \left(2\pi\left (\frac{B_f}{B}-\delta \right)\right)  ,
\label{eqn:cosFunc}
\end{equation} 

\noindent where the $\delta = 1/2$ and $B_f$ is the SdH frequency. Here, we use the conductivity and conductance terms interchangeably since both parameters are related via a simple geometric factor. Fig.\ref{fig:FigS5} shows the background reduced SdH data and the fit performed for CBM302 at 20 $mK$. In it, we used the moving average of the raw data which was then subtracted to obtain the background reduced SdH oscillations. The fitting function used Eq \ref{eqn:cosFunc} to obtain the SdH frequency which was then used to calculate the electron density $n$, 

\begin{equation}
n = \frac{2eB_{f}}{h},
\label{eqn:density}
\end{equation} 

where $e$ and $h$ are the electron charge and Planck's constant.

\newpage
\subsection {Temperature dependence of electron density}

\begin{figure}[!ht]
      \centering
      \includegraphics[width=121mm]{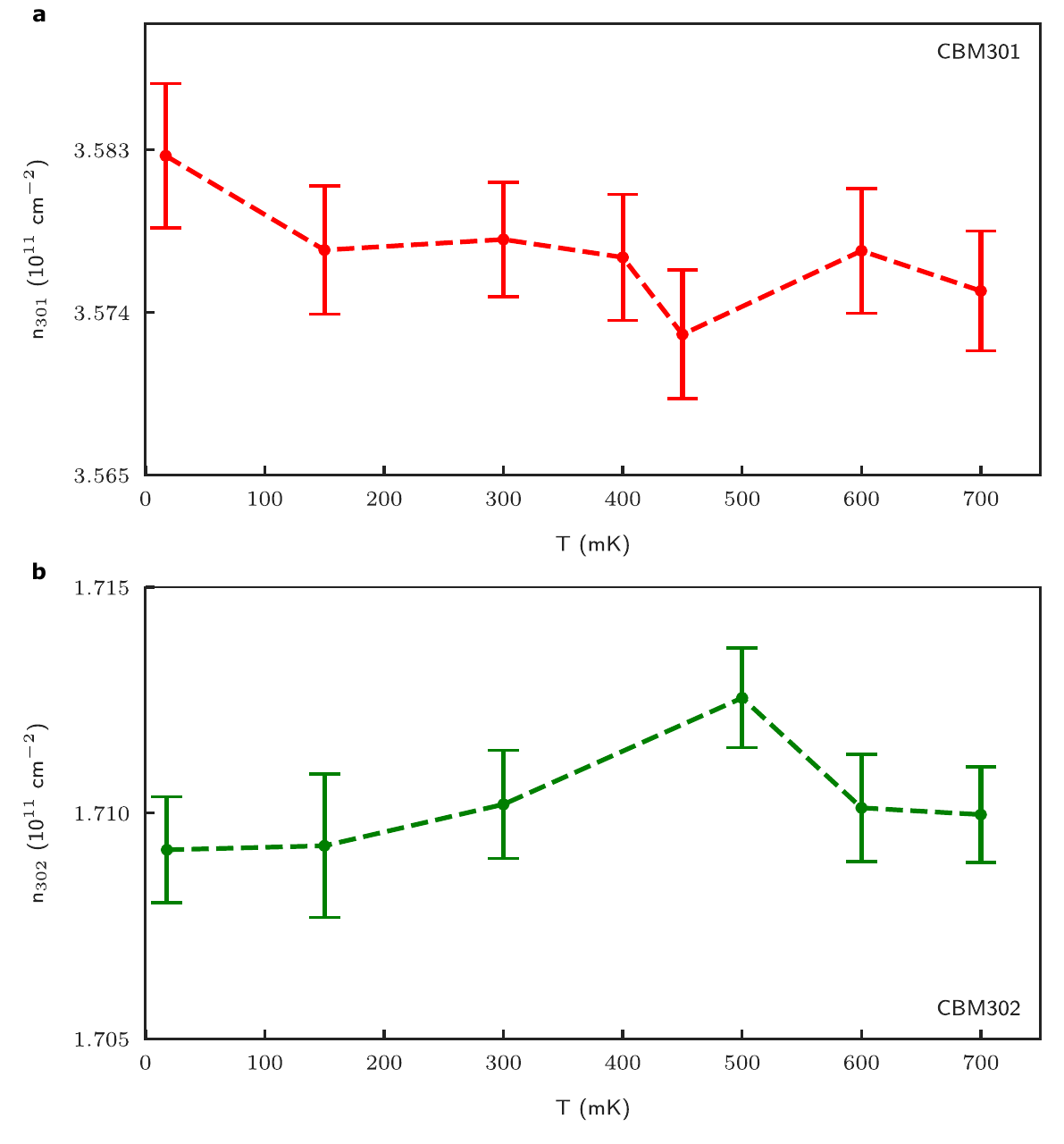}
  \caption{Temperature dependence of electron density for (a) CBM301 and (b) CBM302.}
  \label{fig:FigS6}
\end{figure}

The electron density extracted using the aforementioned procedure at different temperatures ranging from 20 $mK$ to 750 $mK$ for CBM301 and CBM302 are shown in Fig.\ref{fig:FigS6}. The calculated electron density remains constant within $0.5 \%$ in the temperature interval where the anomalous behaviour in electronic transport was observed.


\newpage
\section {Data consistency and reproducibility}
\begin{figure}[!ht]
      \centering
      \includegraphics[width=121mm]{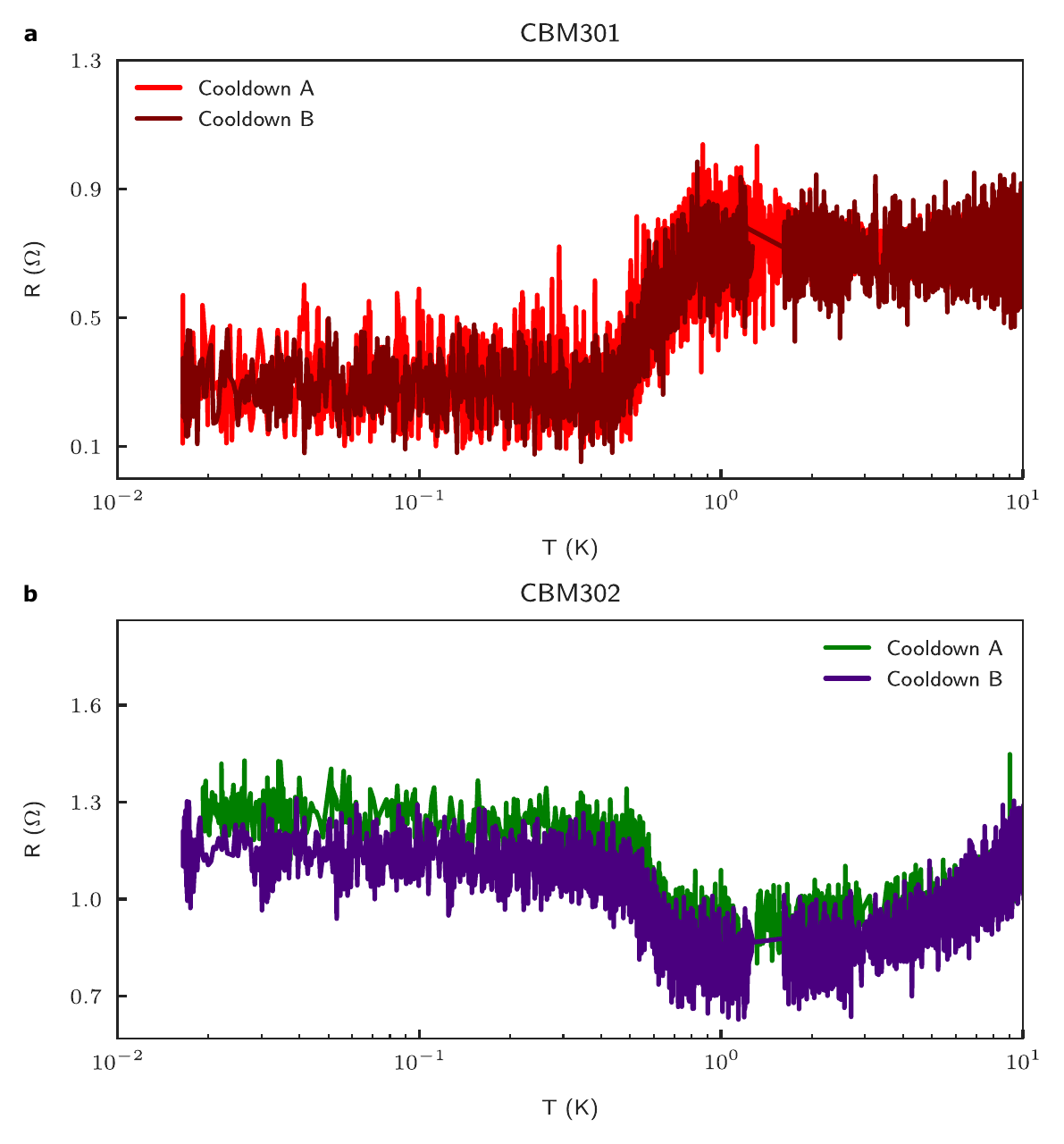}
  \caption{Temperature dependence of resistance for (a) CBM301 (b) and CBM302 measured during two different cooldown.}
  \label{fig:FigS7}
\end{figure}

The anomalous behavior in resistance observed in CBM301 and CBM302 has been reproduced during several cooldowns. Fig.\ref{fig:FigS7} shows the data obtained for two different cooldowns in the same dilution refrigerator system using experimental setup A for both Corbino samples. In addition to the data shown in Fig.\ref{fig:FigS7}, the results have been reproduced several times using both experimental setups in different cooldowns, and also in two different dilution refrigerators with distinctive electronics.

\newpage
\section {Transport measurements at higher temperature}
\begin{figure}[!ht]
      \centering
      \includegraphics[width=121mm]{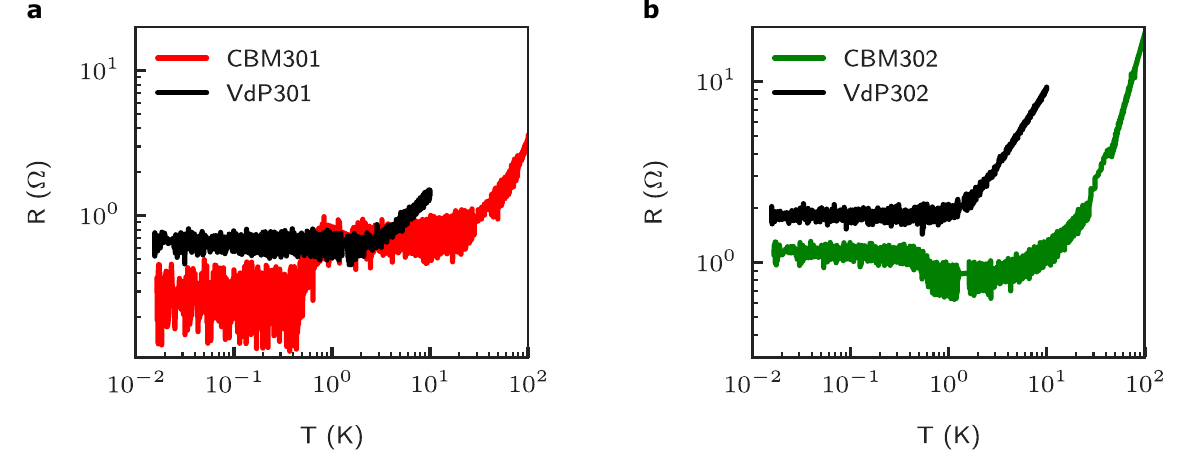}
  \caption{Temperature dependence of resistance up to 100 $K$ for (a) CBM301, VdP301 and (b) CBM302, VdP302.}
  \label{fig:FigS8}
\end{figure}

Electronic transport measurements were also performed at higher temperatures than shown in the main text,  and they are shown in Fig.\ref{fig:FigS8}. As expected, the monotonic increase in resistance with increasing temperature for VdP301 and VdP302 is observed. Note the increase in resistance observed at $\sim$400 $mK$ for CBM301 is followed by a nearly-constant resistance value over a wide range of temperature up to $\sim$10 K,  and subsequently a monotonic increase in resistance is observed at higher temperatures.

\end{document}